\documentstyle[twoside,fleqn,espcrc2]{article}

\begin{document}

\author{Remo Garattini\thanks{%
Talk given at the {\it Second Meeting on Constrained Dynamics and Quantum
Gravity, }Santa Margherita Ligure, September 17-21, 1996, to appear in the
Proceedings.} \\
Facolt\`{a} di Ingegneria, Universit\`{a} di Bergamo\\
Viale Marconi, 5, 24044 Dalmine (Bergamo) Italy\\
E-mail:Garattini@mi.infn.it}
\title{Energy Computation in Wormhole Background with the Wheeler-DeWitt Operators}

\begin{abstract}
We investigate the possibility of computing energy by means of operators
associated to the {\it Wheeler-DeWitt} equation. By choosing three
dimensional wormholes as a framework, we apply such calculation scheme to
the black hole pair creation. We compare our results with the recent ones
appeared in the literature.
\end{abstract}

\maketitle

\section{Introduction}

In recent years, a growing attention is devoted to the subject of wormholes
in its various aspects. Here we would like to use such objects together with
a variational calculation to probe the quantum gravitational vacuum. As an
application, we will consider the pair creation problem for neutral black
holes, whose rate is determined by the following formula

\begin{equation}
P\simeq \left| e^{-I_{cl}}\right| ^{2}\simeq \left| e^{-\left( \Delta
E\right) \left( \Delta t\right) }\right|  \label{i1}
\end{equation}

The sourceless Einstein's equations, with and without cosmological constant,
select two types of spherical symmetric solutions representing a hole in
spacetime:

\begin{itemize}
\item  the Schwarzschild solution (S)

\item  the Schwarzschild-deSitter solution (SdS).
\end{itemize}

According to \cite{MTW}, we shall call such solutions wormholes. Although
the exclusion of matter fields is far from a realistic picture of quantum
gravity, we use such a representation to study quantum effects in the very
early universe. In static coordinates these solutions assume the form

\begin{equation}
ds_{{}}^{2}=\pm F\left( r\right) dt^{2}+F^{-1}\left( r\right)
dr^{2}+r^{2}d\Omega ^{2},  \label{i2}
\end{equation}

where ``$+$'' is for the Euclidean signature while ``$-$'' is for the
Lorentzian one. The function $F\left( r\right) $ is such that

\begin{equation}
\left\{ 
\begin{array}{ll}
1-\frac{2MG}{r} & {(S)} \\ 
1-\frac{2MG}{r}-\frac{1}{2}\Lambda r^{2} & {(SdS)}
\end{array}
\right. ,
\end{equation}

$M$ is the parameter describing the mass of the wormhole, $G$ is the
gravitational constant, $\Lambda $ is the positive cosmological constant and 
$d\Omega ^{2}=d\theta ^{2}+\sin ^{2}\theta d\phi ^{2}$. By observing that,
the constant ``{\it time}'' section of $\left({\ref{i2}}\right) $ is
represented by a three dimensional wormhole which maintains metric and
topology for both signatures, we are led to consider $3D$ wormholes as
preferred frame for probing the vacuum. Nevertheless, for our purposes, we
shall make the choice of completing the spacetime in the compact Euclidean
direction. More details on this subject can be found in Ref. \cite{Remo2}

\section{Instantons and Wormholes}

We report here the analogies between gravitational instantons and wormholes
related to the metrics (topologies) under examination.

\subsection{Instantons}

\label{p1}The instantons associated to the problem under examination are 
\[
\begin{array}{lll}
a) & {Nariai }\left( N\right) & {\bf S}^{2}\times {\bf S}^{2} \\ 
b) & {\ deSitter }\left( dS\right) & {\bf S}^{4} \\ 
c) & {Gibbons-Hawking }\left( GS\right) & {\bf R}^{2}\times {\bf S}^{2} \\ 
d) & {Flat Space }\left( FS\right) & {\bf R}^{3}\times {\bf S}^{1}.
\end{array}
\]

In particular in a) and in b) no boundary terms are needed because the
corresponding topologies are compact. Because the (SdS) metric tends
asymptotically to the (dS) metric, we can consider the latter one as a
reference background. Moreover, for regularity reasons we refer to the (N)
instanton that is the extreme (SdS) instanton \cite{Nariai}. For these two
instantons (topologies) the value of the action, for $\Lambda >0$, is

\begin{equation}
\begin{array}{l}
-\frac{3\pi }{G\Lambda }{\qquad (dS)} \\ 
-\frac{2\pi }{G\Lambda }{\qquad (N)}
\end{array}
\end{equation}

In c) and in d) the action needs a boundary term, otherwise the path
integral is meaningless. Here $\Lambda =0$ and

\begin{equation}
I=-\frac{1}{8\pi G}\int\limits_{{\cal \partial M}}d^{3}x\sqrt{h}\left[
K\right] ,
\end{equation}

where $\left[ K\right] $ is the difference in the trace of the second
fundamental form of ${\cal \partial M}$ in the metric $g$ and the metric $%
\eta $ referred to the flat space. Then $I=4\pi GM^{2},$where we have used
the fact that the Euclidean ``time'' is periodic with period $8\pi GM$ and
the fact that the hypersurface is bounded by the surface $r=r_{0}$.

\subsection{Wormholes}

Turning to the wormhole sector, the scalar curvature appearing in the action
is better described in terms of {\it lapse} and {\it shift} variables. The
contribution for these metrics comes only from a boundary term. This is a
consequence of the fact that we have opened the hypersurfaces, i.e. the
manifold is no more a compact object. Then, the relation between the four
dimensional spacetime and the three dimensional space plus one compact time,
with and without cosmological constant, is

\begin{equation}
^{\left( 4\right) }R=\frac{2}{N}\nabla ^{2}N
\end{equation}

To check that the action contribution is the same of the Sec.$\left( {\ref
{p1}}\right) $, we have to integrate over the period the boundary term. This
process gives the relation between the four compact dimension and the three
plus one compact time dimensions.

\begin{description}
\item[a)]  (dS) vs. ${\bf S}^{4}$ topology.

The line element is of the form of the eq $\left( {\ref{i1}}\right) $. By
identifying $N^{2}\left( r\right) =F\left( r\right) =1-\frac{\Lambda }{3}%
r^{2}$, the action becomes 
\[
I=-\frac{1}{8\pi G}\int d\tau \int d^{3}x\left[ \partial _{i}\left( \sqrt{%
^{3}g}g^{ij}\partial _{j}N\right) \right] 
\]

\begin{equation}
=-\frac{\Lambda }{6G}\int d\tau r^{3}|_{=\sqrt{\frac{3}{\Lambda }}}=-\frac{%
3\pi }{\Lambda G}.  \label{c6}
\end{equation}

\item[b)]  (N) vs. ${\bf S}^{2}\times $ ${\bf S}^{2}$ topology.

By identifying $N^{2}\left( r\right) =F\left( r\right) =1-\Lambda r^{2}$, we
obtain

\begin{equation}
I=-\frac{1}{8\pi G}\left( \frac{2\pi }{\sqrt{\Lambda }}\right) \left( 4\pi 
\frac{1}{\Lambda }\right) \int d\tau =-\frac{2\pi }{G\Lambda }.  \label{c8}
\end{equation}

\item[c)]  (S) vs. ${\bf R}^{2}\times $ ${\bf S}^{2}$ topology.

By identifying $N^{2}\left( r\right) =F\left( r\right) =1-\frac{2M}{r}$, one
obtains directly the same result of the instanton because in that case the
only contribution comes from the boundary.
\end{description}

\section{Calculation Scheme and Application to the Black Hole Pair Creation}

From the previous section we have seen how to recover instanton results
starting from $3D$ wormhole at the classical level. On these grounds we
approach quantum gravitational effects:

\begin{itemize}
\item  Our framework will be a variational calculation applied to gaussian
wave functionals $\Psi \left\{ g_{ij}^{{}}\left( x\right) \right\} $ . The
central point is the evaluation of

\begin{equation}
\frac{\langle \Psi \left| H\right| \Psi \rangle }{\langle \Psi |\Psi \rangle 
},  \label{b1}
\end{equation}

where $H$ is the total Hamiltonian which contains two classical constraints

\begin{equation}
\left\{ 
\begin{array}{l}
{\cal H}=0 \\ 
{\cal H}_{i}=0
\end{array}
\right. .
\end{equation}

The operator version of the first of the two previous constraints is the 
{\it Wheeler-DeWitt} equation. It is immediate to see that the calculation
of the quantity expressed in eq.$\left( {\ref{b1}}\right) $ is ill posed.
However, we can proceed backward by evaluating the average of $H$ in an
enlarged {\it nonphysical }space, postponing the projection on the physical
Hilbert space by imposing the {\it Wheeler-DeWitt} equation.

\item  We will consider small perturbations with respect to the constant ``%
{\it time}'' section (S) and (N) metric that we denote as $\tilde{g}_{ij}$.
Then the perturbations will be in the {\it three dimensional} space and

\begin{equation}
g_{ij}=\tilde{g}_{ij}+h_{ij},
\end{equation}

with

\begin{equation}
\langle \Psi \left| g_{ij}\right| \Psi \rangle =\tilde{g}_{ij}.
\end{equation}

\item  In this context, the {\it shift }function is zero, then only ${\cal H}
$ is non vanishing in the{\it \ enlarged space.} We will denote this
operator ${\cal H}_{WDW}$.{\it \ }The expansion of ${\cal H}_{WDW}$ to one
loop gives rise to the second order differential operator acting only on 
{\it gravitons} $\left( {TTsector}\right) $.

\item  The analysis of the discrete spectrum for

\begin{equation}
\left( -\triangle \delta _{j}^{a_{{}}^{{}}}+2R_{j}^{a}\right)
h_{a}^{i}=-E^{2}h_{j}^{i_{{}}^{{}}}
\end{equation}

gives one negative squared eigenvalue for both metrics, where $R_{j}^{a}$ is
the Ricci tensor in $3D$ and $\triangle $ is the Laplacian in curved space.

\begin{equation}
\begin{array}{ll}
-.24\left( GM\right) ^{-2} & {(GH)} \\ 
-2\Lambda  & {(N).}
\end{array}
\end{equation}

\item  Negative squared eigenvalues appear for the graviton sector in $4D$
by means of the expansion of the Euclidean action at one loop \cite
{Gross,Perry,Young}

\begin{equation}
\begin{array}{ll}
-.19\left( GM\right) ^{-2} & {(GH)} \\ 
-2\Lambda  & {(N).}
\end{array}
\end{equation}

\item  Presence of negative modes in three dimensions plus one {\it compact}
dimension is guaranteed by the same analysis in a Kaluza-Klein manifold with
one dimension suppressed.\cite{Witten}

\item  Negative modes {\it imply} the decay from the {\it false vacuum }to
the {\it (possible) true vacuum.} The decay probability per unit time and
unit volume is defined, at least semiclassically as

\begin{equation}
\Gamma =A\exp \left( -I_{cl}\right) ,
\end{equation}

where $A$ is the prefactor coming from the saddle point evaluation and $%
I_{cl}$ is the classical part of the action.

\item  Neglecting the prefactor \cite{Coleman}, the approximate value of the
probability of nucleating a black hole in the different instantons
(topologies) is

\begin{equation}
\Gamma \sim \left\{ 
\begin{array}{ll}
\exp -4\pi M^{2}G & {(GH)} \\ 
\exp -\frac{\pi }{\Lambda G}{\label{eu}} & {(N)}
\end{array}
\right. 
\end{equation}
\end{itemize}

In any case the ``{\it hot}'' space cannot describe the ground state,
therefore a topology change comes into play and a black hole nucleation can
be realized when we consider the {\it hot flat space}, while black hole pair
creation is the mechanism related to the {\it hot de Sitter space. }It is
immediate to recognize that eq. $\left(\ref{eu}\right) $ represents the
decay rate calculated with the no-boundary prescription of Hartle-Hawking.
In fact, following Ref. \cite{Bousso-Hawking}, we define

\begin{equation}  \label{de}
\Gamma =\frac{P_{SdS}}{P_{de Sitter}}=\exp -\frac{\pi }{G\Lambda }.
\end{equation}

However in eq. $\left(\ref{i1}\right) $, the form of the decay probability
can be expressed in terms of the wavefunction solving the WDW equation. For
this purpose we need to project out the non-physical states from the initial
enlarged space. A possibility to do this is by means of the following choice 
\cite{Remo1}

\begin{equation}  \label{wf}
\Psi \left[ g_{ij}\right] =\int\limits_{\gamma }dN\ e^{-N\omega }\Phi \left[
g_{ij}\right] .
\end{equation}

However, since we have introduced boundary terms the previous equation has
to be modified with

\begin{equation}
\Psi \left[ g_{ij}\right] =\exp \left( I_{b.t.}\right) \int\limits_\gamma
dN\ e^{-N\omega }\Phi \left[ g_{ij}\right] ,
\end{equation}

where

\begin{equation}
I_{b.t.}=\frac 1{8\pi }\int_{\tau =0}d^3x\sqrt{^3g}K.
\end{equation}

The only difference with the Hartle-Hawking wave function is that $\Phi
\left[ g_{ij}\right] $ is a trial wave functional of the gaussian type \cite
{Remo}. By repeating the same procedure of cutting half of the instanton, we
recover the results that lead to eq. $\left(\ref{de}\right) .$ The same
approach can be applied to the (GH) instanton \cite{Frolov}.

\section{Conclusions and Outlooks}

\begin{itemize}
\item  Wormholes without matter fields could be significant in the very
early universe

\item  The probability of decay is relevant when $\Lambda \sim 1$ ( in
Planck's units)

\item  Nucleation happens in the hot space.
\end{itemize}

\medskip

We need to investigate:

\begin{itemize}
\item  The cold space, i.e. $T=0.$

\item  Higher order corrections

\item  The conformal factor

\item  Matter field contribution

\item  Multi-wormholes$\stackrel{?}{\longleftrightarrow }$ spacetime foam.
\end{itemize}

\section{Acknowledgments}

I would like to thank Prof. V. de Alfaro and M. Cavagli\`{a} who gave me the
opportunity of presenting this talk.

\end{document}